\documentclass[twocolumn]{aastex631}  
\usepackage{amsmath}
\usepackage{graphicx}
\usepackage{threeparttable}
\usepackage{txfonts}
\hypersetup{allcolors=blue}

\newcommand{\msun}{{$M_{\odot}$}}

\newcommand{\ledd}{$L_{\rm Edd}$}

\newcommand{\gtsima}{$\; \buildrel > \over \sim \;$}
\newcommand{\ltsima}{$\; \buildrel < \over \sim \;$}
\newcommand{\prosima}{$\; \buildrel \propto \over \sim \;$}
\newcommand{\gsim}{\lower.5ex\hbox{\consistegtsima}}
\newcommand{\lsim}{\lower.5ex\hbox{\ltsima}}
\newcommand{\simgt}{\lower.5ex\hbox{\gtsima}}
\newcommand{\simlt}{\lower.5ex\hbox{\ltsima}}
\newcommand{\simpr}{\lower.5ex\hbox{\prosima}}

\newcommand{\lx}{$L_{\rm X}$}

\newcommand\iona[2]{#1$\;${\scshape{#2}}}

\begin{document}  

\title{Gone with the Wind: JWST-MIRI Unveils a Strong Outflow from the Quiescent Stellar-Mass Black Hole A0620-00}

\author[0009-0000-8524-8344]{Zihao Zuo}
\affiliation{Department of Astronomy, University of Michigan, 1085 S University, Ann Arbor, MI 48109, USA}
\author[0000-0001-7255-3251]{Gabriele Cugno}
\affiliation{
Department of Astrophysics, University of Zurich. 
Winterthurerstrasse 190, 8057 Zurich, Switzerland}
\author[0000-0003-3503-3446]{Joseph Michail}
\altaffiliation{NSF Astronomy \& Astrophysics Postdoctoral Fellow}
\affiliation{Center for Astrophysics, Harvard \& Smithsonian, 60 Garden Street, Cambridge, MA, USA 02138}
\author[0000-0001-5802-6041]{Elena Gallo}
\affiliation{Department of Astronomy, University of Michigan, 1085 S University, Ann Arbor, MI 48109, USA}
\author[0000-0002-3500-631X]{David M. Russell}
\affiliation{Center for Astrophysics and Space Science (CASS), New York University Abu Dhabi, PO Box 129188, Abu Dhabi, UAE}
\author[0000-0002-7092-0326]{Richard M. Plotkin}
\affiliation{Physics Department, University of Nevada, Reno, 1664 N. Virginia St, Reno, NV, 89557, USA}
\affiliation{Nevada Center for Astrophysics, University of Nevada, Las Vegas, NV 89154, USA}
\author[0000-0002-4436-6923]{Fan Zou}
\affiliation{Department of Astronomy, University of Michigan, 1085 S University, Ann Arbor, MI 48109, USA}
\author[0000-0003-1285-4057]{M. Cristina Baglio}
\affiliation{INAF-Osservatorio Astronomico di Brera, Via Bianchi 46, I-23807 Merate (LC), Italy}
\author[0000-0002-0752-3301]{Piergiorgio Casella}
\affiliation{INAF-Osservatorio Astronomico di Roma, Via Frascati 33, I-00076 Monte Porzio Catone (RM), Italy}
\author[0009-0009-0079-2419]{Fraser J. Cowie}
\affiliation{Astrophysics, Department of Physics, University of Oxford, Keble Road, Oxford OX1 3RH, UK}
\author[0000-0002-5654-2744]{Rob Fender}
\affiliation{Astrophysics, Department of Physics, University of Oxford, Keble Road, Oxford OX1 3RH, UK}
\affiliation{Department of Astronomy, University of Cape Town, Private Bag X3, Rondebosch 7701, South Africa}
\author[0000-0003-3105-2615]{Poshak Gandhi}
\affiliation{School of Physics \& Astronomy, University of Southampton, Southampton
SO17 1BJ, UK}
\author[0000-0001-9564-0876]{Sera Markoff}
\affiliation{Anton Pannekoek Institute for Astronomy/GRAPPA, University of Amsterdam, Science Park 904, 1098 XH Amsterdam, NL}
\author[0000-0002-1481-1870]{Federico Vincentelli}
\affiliation{INAF - Istituto di Astrofisica e Planetologia Spaziali, Via del Fosso del Cavaliere 100, I-00133 Roma, Italy}
\author[0000-0003-3352-2334]{Fraser Lewis}
\affiliation{Faulkes Telescope Project, School of Physics and Astronomy, Cardiff University, The Parade, Cardiff, CF24 3AA Wales, UK}
\affiliation{The Schools’ Observatory, Astrophysics Research Institute, Liverpool John Moores University, 146 Brownlow Hill, Liverpool L3 5RF, UK}

\author[0000-0003-2869-7682]{Jon M. Miller}
\affiliation{Department of Astronomy, University of Michigan, 1085 S University, Ann Arbor, MI 48109, USA}
\author[0000-0003-3124-2814]{James C.A. Miller-Jones}
\affiliation{International Centre for Radio Astronomy Research, Curtin University, GPO Box U1987, Perth, WA 6845, Australia}

\author[0000-0002-5767-7253]{Alexandra Veledina}
\affiliation{Department of Physics and Astronomy, FI-20014 University of Turku, Finland}
\affiliation{Nordita, KTH Royal Institute of Technology and Stockholm
University, Hannes Alfv\'ens v\"ag 12, SE-10691 Stockholm, Sweden}

\begin{abstract}
We present new observations of the black hole X-ray binary A0620$-$00 using the Mid-Infrared Instrument on the James Webb Space Telescope, during a state where the X-ray luminosity is 9 orders of magnitude below Eddington, and coordinated with radio, near-infrared and optical observations.
The goal is to understand the nature of the excess mid-infrared (MIR) emission originally detected by Spitzer red-ward of 8~$\mu$m. 
The stellar-subtracted MIR spectrum is well-modeled by a power law with a spectral index of $\alpha=0.72\pm0.01$, where the flux density scales with frequency as $F_\nu \propto \nu^{\alpha}$. The spectral characteristics, along with rapid variability--a 40\% flux flare at 15$\mu$m and 25\% achromatic variability in the 5-12$\mu$m range--rule out a circumbinary disk as the source of the MIR excess. The Low Resolution Spectrometer reveals a prominent emission feature at 7.5$\mu$m, resulting from the blend of three hydrogen recombination lines. While the contribution from partially self-absorbed synchrotron radiation cannot be ruled out, we argue that thermal bremsstrahlung from a warm (a few $10^4$ K) wind accounts for the MIR excess; the same outflow is responsible for the emission lines. The inferred mass outflow rate indicates that the system's low luminosity is due to a substantial fraction of the mass supplied by the donor star being expelled through a wind rather than accreted onto the black hole.
\end{abstract}

\section{Introduction}
Black hole X-ray binaries (XRBs) spend the majority of their lifetimes in a low-luminosity, ``quiescent" state and occasionally undergo dramatic outbursts lasting weeks to months, likely triggered by viscous-thermal disk instabilities \citep[e.g.,][]{Lasota2001}. During these outbursts, their luminosity increases by orders of magnitude across all wavelengths. Despite long outburst recurrence times (years to tens of years), our understanding of the accretion process in stellar-mass black holes is largely based on systems in outburst when the X-ray luminosity approaches the Eddington limit \citep[see, e.g.,][for reviews on X-ray states and modeling accretion in black hole XRBs]{Remillard2006, Done2007, Belloni2011}. Luminous, hard X-ray states are typically associated with persistent radio emission with a flat or slightly inverted spectrum ($F_{\nu} \propto \nu^{\alpha}$, with $\alpha \approx 0-0.5$) \citep{Fender2001}, arising from a partially self-absorbed jet. This jet becomes progressively more transparent at longer wavelengths as it propagates toward larger distances from its base \citep[cf.][]{Blandford1979}. The flat-inverted jet spectrum extends up to the near-IR band, where it becomes optically thin \citep{Corbel2002,Gandhi2011,Russell2013}. When sufficiently bright, the radio emission is resolved into a collimated jet on milliarcsecond scales \citep{Dhawan2000, Stirling2001, Wood2024}.

The observational properties of quiescent black hole XRBs, with $L_X \ll 10^{-4} L_{\rm Edd}$, differ significantly from those during outbursts. The accretion flow becomes highly radiatively inefficient, and the relative contributions of inflow versus outflow to the overall spectral energy distribution (SED) are poorly constrained \citep[e.g.,][]{YuanNarayan2014}. More generally, the modeling of accretion and jet production at low accretion rates is still a highly debated topic within the high-energy astrophysics community.
\par
Relativistic jets are thought to persist in quiescence (\citealt{Plotkin2021}, and references therein; see \citealt{Fender2009} for a comprehensive review of the radio properties of black hole XRBs across different X-ray states). A handful of quiescent black hole XRBs have indeed been detected as flat or inverted spectrum, compact radio sources \citep[e.g.,][]{dePolo2022}. However, the low, quiescent radio flux densities make it difficult, if not impossible, to resolve a collimated jet on milliarcsecond scales directly.
\par
The wavelength at which the jet synchrotron emission becomes optically thin (referred to as jet break) in quiescent systems remains uncertain. The location of the break is crucial for placing limits on the potential jet synchrotron contributions at ultraviolet (UV) and X-ray energies. Repeated multiwavelength studies of luminous, hard state black hole XRBs have consistently placed this thick-to-thin jet break frequency in the near-IR band \citep{Gandhi2011,Russell2013}, but a {\it direct} measurement for the quiescent population is lacking.
\par
If the partially self-absorbed synchrotron emission of quiescent systems also extends into the IR band, it would imply that the jet power vastly exceeds that of the X-ray emitting accretion flow \citep{Fender2003, Gallo2006, Gallo2007}. Confirmation of this observationally would challenge the classical view that accreting black holes emit the majority of the locally dissipated accretion power within the inflow. 
Several studies suggest that the break could occur at longer wavelengths in quiescence compared to the hard state \citep{Shabaz2013}, or at the very least, be variable \citep{Plotkin2016, Russell2018}, as also inferred for the more luminous hard state \citep{Gandhi2011}.%
\begin{table*}[t!]
\centering{
\caption{JWST MIRI Observation Summary}
\def\arraystretch{1.15}
\begin{tabular}{lllllllllllc}\hline\hline
Mode/Filter & $\lambda_\mathrm{pivot}$ & W$_\mathrm{eff}$\tablenotemark{\scriptsize a} & Readout/Subarray\tablenotemark{\scriptsize b} & $N_\mathrm{gr}$ & $N_\mathrm{int}$ & $N_\mathrm{dither}$ & $t_\mathrm{tot}$ & Start Time & S/N &Flux (mJy)\\
 & ($\mu$m)  & ($\mu$m) &  & & & & (s) & (UTC) \\ \hline
Imager/F1500W  & 15.06 & 2.92 & FASTR1/FULL & 6 & 1 & 4 & 66.6  & 2024-03-17 08:16:52 & 93 & $0.193\pm0.004$ \\    
Imager/F1800W  & 17.98 & 2.95 & FASTR1/FULL & 6 & 1 & 4 & 66.6  & 2024-03-17 08:24:52 & 47 & $0.126\pm0.003$ \\   
Imager/F2100W  & 20.79 & 4.58 & FASTR1/FULL & 6 & 1 & 4 & 66.6  & 2024-03-17 08:33:03 & 36 & $0.112\pm0.002$ \\ 
Imager/F2550W  & 25.36 & 3.67 & FASTR1/FULL & 10 & 7 & 4 & 843.6 & 2024-03-17 08:41:53 &  13 & $0.12\pm0.04$\\ 
LRS         &  &  & FASTR1/FULL & 23& 5 & 2 & 660.4 & 2024-03-17 09:17:22 & $-$ & 0.161 -- 0.481 \\\hline\hline
\end{tabular}\\\vspace{0.2cm}
}
\tablenotetext{\scriptsize a}{Filter bandwidth, defined as the integral of the normalized transmission curve ({\url{https://jwst-docs.stsci.edu/jwst-near-infrared-camera/nircam-instrumentation/nircam-filters}}).}
\tablenotetext{\scriptsize b}{FULL has a group time of $t_\mathrm{gr}$ of 2.775 s. Each integration length was $t_\mathrm{gr} \times N_\mathrm{gr}$.}
\label{tab:observations}
\end{table*}
%
Possible direct evidence indicating that the jet's synchrotron emission might contribute to the mid-infrared (MIR) band in highly sub-Eddington systems comes from Spitzer Space Telescope observations of the black hole XRBs A0620-00 and XTE~J1118+480 \citep{Gallo2007}. In both systems, Spitzer detected excess MIR emission relative to the tail of the donor star's photosphere at 8 $\mu$m and, in the case of A0620-00, at 24 $\mu$m. The contribution from the accretion disk is thought to be negligible at these wavelengths. However, the large error bars at 24 $\mu$m allow for two radically different interpretations of this excess: either partially self-absorbed synchrotron emission with a slightly inverted spectrum, or blackbody emission from circumbinary material reprocessing the donor star's light \citep{Muno2006}. 
The inference of a circumbinary nature for the MIR excess arises from the fact that, for both systems, the temperatures and normalizations of the best-fit blackbody spectra ($393 \pm 83$ K for A0620-00 and $754 \pm 140$ K for XTE~J1118+108) imply radii approximately twice the orbital separation (according to theoretical predictions \citep{Taam2001, Dubus2001, Taam2003}, the temperature of a circumbinary disk depends on its inner radius, and the temperature and size of the donor star). In this scenario, the jet spectrum arguably becomes optically thin somewhere between the millimeter and MIR wavelengths.
\par
From an evolutionary perspective, a circumbinary disk may form at early evolutionary stages, such as during the binary common envelope phase. Alternatively, it could be fed or replenished by mass outflows from the donor star or the outer accretion disk. Possible indirect evidence for outflowing matter has been reported in a handful of cataclysmic variables, inferred from single-peaked emission lines with narrow widths \citep[e.g.,][]{Hellier2000}. Strongly ionized, mass-loaded accretion disk winds are routinely detected in black hole XRBs during outbursts through X-ray absorption spectroscopy \citep[e.g.,][]{Ponti2012, Miller2015}, as well as in the optical/IR \citep[e.g.,][]{Munoz-Darias2016}. However, these systems are all too active to enable direct searches in the MIR.\\
%
\begin{figure*}[t!]
\centering
\includegraphics[width=0.9\linewidth,angle=0]{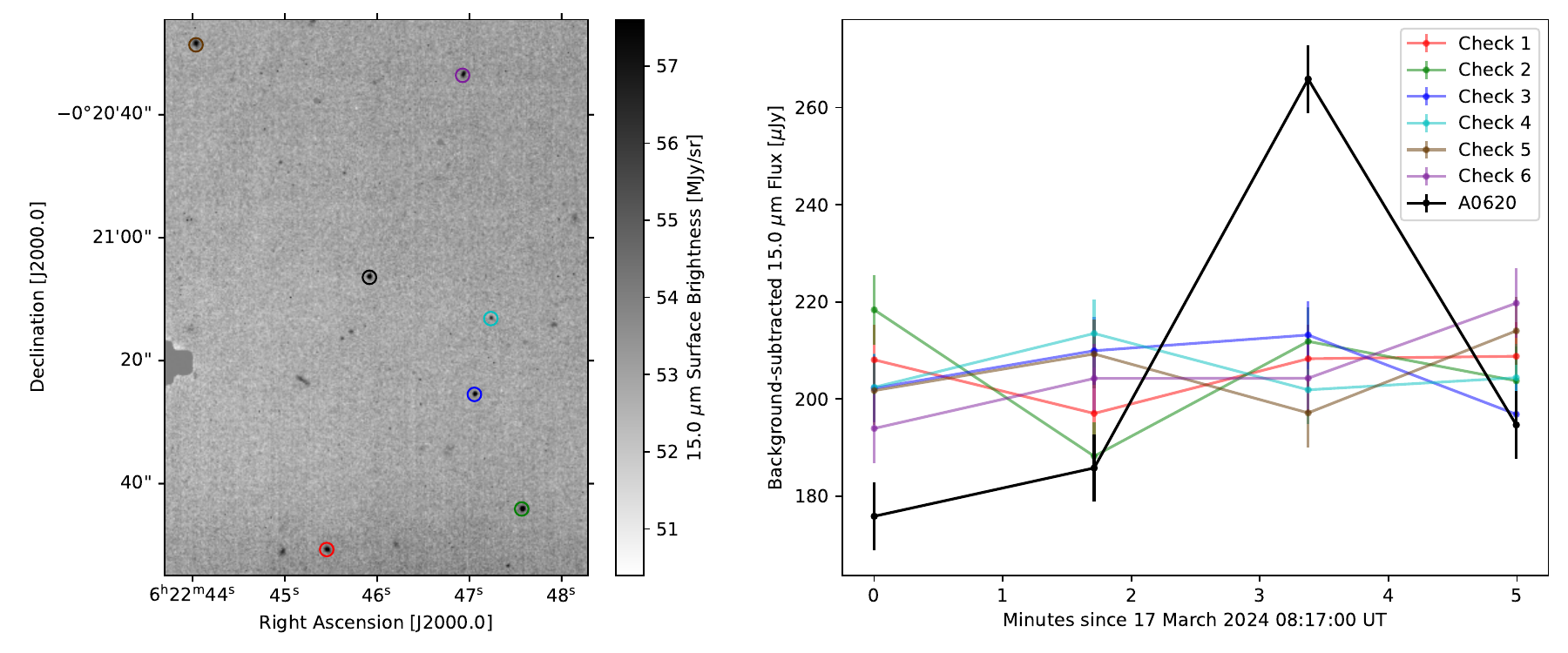}
\caption{\textit{Left}: A map of the A0620 field in the F1500W filter during one of the dithers. A0620 (black circle) is near the center of the detector in this image. We detect six nearby sources visible in all four dithers, which we use to check for residual instrumental calibration errors. \textit{Right}: Background-subtracted aperture photometry of all seven sources (line colors correspond to colored circles in left panel) in the field during the four-dither sequence. The check source light curves have been shifted to match the average flux of A0620. We detect no correlated changes in the check sources corresponding to the measured $\sim$40\% variability in A0620.}
\label{fig:flare}
\end{figure*}
%
%

With its exquisite sensitivity in the MIR, The Mid-Infrared Instrument (MIRI) onboard the James Webb Space Telescope (JWST) offers the opportunity to solve the long-standing puzzle about the nature of the MIR excess of quiescent Galactic black hole XRBs. At \lx$\simeq 10^{-9}$ \ledd, and $1.5^{+0.3}_{-0.2}$ kpc distance \citep{Zhao2023}, A0620-00 (A0620 hereafter) provides us with the ideal testbed for competing models. In this Paper, we report on JWST MIRI observations of A0620, obtained in March 2024. In concert with closely-spaced radio and near-IR-optical observations, these data enable us to place new stringent constraints on the properties of highly sub-Eddington stellar-mass black holes.\par
Throughout, system's parameters such as the circularization radius and or accretion disk outer radius are estimated assuming a black hole mass of 6.6 \msun, and period of 7.75 hours \citep{Cantrell2010}.   

\section{Observations, Data Reduction, and Analysis}

A0620 was observed by the JWST MIRI on 2024 March 17, as part of the General Observer (GO) program ID 3832 (PI: Gallo). A comprehensive multi-wavelength campaign supported the JWST effort. Nearly simultaneous observations were conducted in the radio using the Very Large Array (VLA; PI: Plotkin) and MeerKAT (PI: Fender). Optical photometric data were acquired through the Las Cumbres Observatory network (LCO; PI: Russell) on six occasions within 2.5 days before and after the JWST observation. Near-infrared data were collected within a few hours of the JWST observations using the Very Large Telescope (VLT) High Acuity Wide field K-band Imager (HAWK-I; PI: Casella). Additionally, optical/IR photometric data were obtained with the Rapid Eye Mount (REM) telescope within a few hours of the JWST observations, and optical polarimetry was acquired with the VLT on 2024 March 19 (PI: Baglio). These observations are complemented by long-term optical monitoring of A0620 using the Faulkes Telescopes (PI: Lewis). In this section, all magnitudes are reported using the AB system, except for the $H$-band magnitudes, which are given in the Vega system. 
\subsection{JWST MIRI} 
JWST MIRI observed A0620-00 using the Imager in the F1500W, F1800W, F2100W, and F2550W bands, as well as the Low Resolution Spectroscopy (LRS) mode \citep{Wright2023}. Observations took place between UTC 2024-03-17 08:16:52 and 09:29:45. A detailed description of the observational setup is provided in Table~\ref{tab:observations}.
\par
For the Imager data, we download the stage 3 data products (\texttt{i2d}) from MAST. The background in each filter is estimated using the \texttt{Background2D} class from the Python package \texttt{photutils}, which employs sigma clipping (\texttt{sigma=3}). The \texttt{MMMBackground} background estimator, using the form $(3\times\rm{median} - 2\times\rm{mean})$, is subtracted from the image. The flux of A0620-00 is estimated within an aperture radius that represents 80\% of the encircled energy for each filter; corresponding aperture corrections are subsequently applied (aperture radii and corrections are taken from the \texttt{apercorr\_0010.fits} file). When the signal-to-noise ratio (SNR) is at least $\approx 30$, photometric uncertainties are primarily due to the MIRI absolute flux calibration, which carries a 2\% uncertainty \citep{Gordon2025}. For F2550W, the signal-to-noise SNR is only 13. To estimate the uncertainties, we used the individual dithers obtained during the observations. Forcing the source location to R.A. = $95.68561^\circ$ and Dec. = $-0.345628^\circ$ (J2000) \citep{Muno2006}, we use the same steps described above to measure the flux in each individual dithers. The uncertainty is calculated as the standard deviation of the individual flux measurements, with the absolute flux calibration uncertainty added in quadrature. 
\par
To reduce the LRS data, we download the stage 1 data from MAST and manually apply the \texttt{Spec2Pipeline} and \texttt{Spec3Pipeline} processing steps. We correct outliers in the individual dithers by replacing their values with the mean of the eight surrounding pixels. The spectrum is extracted in two ways: first, by combining both dither images and applying the \texttt{extract\_1d} function on the co-added data; and second, by extracting spectra from each dither individually. For the latter, we use a box aperture of size eight pixels, centered on the signal trace at 5 $\mu$m. By comparing the different extractions, we ensure that any spectral features identified are genuine and not artifacts present in only one of the two exposures. This procedure confirms the absence of artifacts in the spectrum; hence, for the remainder of this manuscript, we will use only the spectrum obtained through the \texttt{jwst} pipeline. Due to the sharp drop in LRS throughput at the spectral region's edges, we restrict our analysis to the 5-12 $\mu$m range. Uncertainties are obtained directly from the pipeline's output.
\par
The results from this analysis are summarized in Table~\ref{tab:observations} and shown in red in Figure \ref{fig:sed}.
\begin{figure}
\includegraphics[trim=0 0 1.25cm 0, clip, width=0.95\linewidth,angle=0]{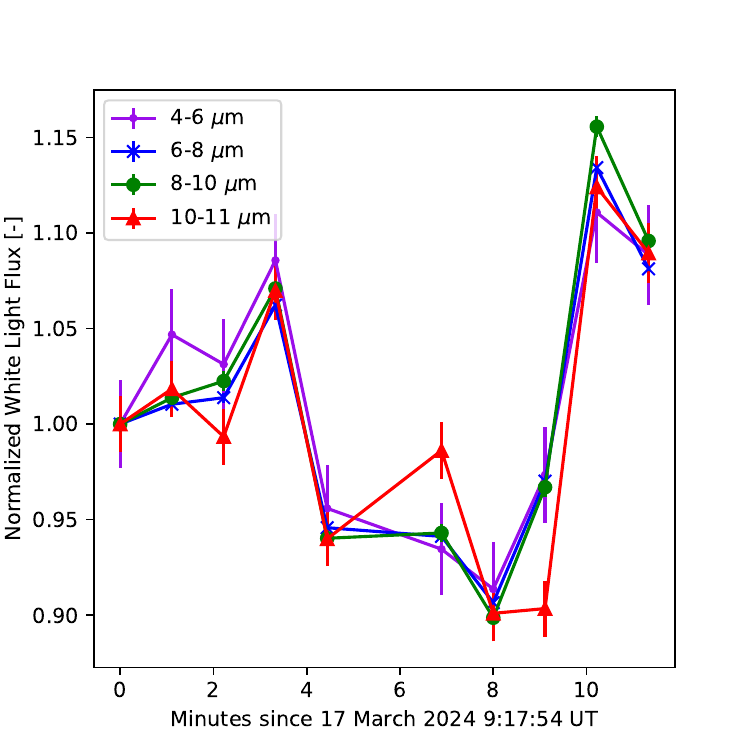}
\caption{Normalized white-light (summed) photometry of A0620 in four LRS wavelength sub-bands during the two-dither sequence. Achromatic variability of $\sim$25\% over the 10-minute sequence is detected in all bands.}
\label{fig:var}
\end{figure}

\subsubsection{MIR Variability}

To search for intrinsic variability of A0620 in the JWST data, we process the Imager and LRS observations through a slightly different pipeline so as not to combine the dithers into a single product. For the F1500W, F1800W, and F2100W, we download the Level 2 \texttt{cal} files for each dither separately. To ensure a uniform background between the four dithers for a single filter, we first process the ensemble through the \texttt{SkyMatch} step in the Level 3 MIRI pipeline. Once complete, we run each dither separately through the remaining steps in the Level 3 pipeline, producing \texttt{i2d} files, giving four measurements of A0620's flux in time for each filter. We do not complete this process for the F2550W data due to the substantial sky background and lower SNR. A similar process of aperture photometry is completed with \texttt{photutils} to obtain the background-subtracted flux of A0620 in each dither. The results for the F1500W filter are shown in Figure \ref{fig:flare}, where we detect an approximate 40\% increase in A0620's 15 $\mu$m flux during the third dither.
To check for systematics, we also determine the background-subtracted light curve for six other sources in the field, which are common to all four dithers and plotted in Figure \ref{fig:flare}. No similar variability is observed in the check source light curves. As a final test, we also measured the average sky background during the four-dither sequence using a background aperture (not shown), finding that it varies by $\lesssim0.5$\%. This indicates that the observed variability of A0620 must be intrinsic. As a result of this detected variability, we adjusted the photometric uncertainty for the F1500W filter to be the standard deviation of the four dither measurements. The same procedure was applied to the other two filters; however, no statistically significant variability was detected in those cases.
\par
For the LRS, we separately process the Level 1 \texttt{rateint} files from MAST for each dither through the Level 2 MIRI spectroscopy pipeline. The first dither is set as the science exposure and the second dither as the background exposure. We then process the second dither oppositely, using the first dither as the background. This procedure produces two \texttt{calint} files, each with five time-resolved, background-subtracted LRS spectra. For a residual background correction in the final time-resolved spectra, we use a rectangular aperture around the positive spectral trace and two background apertures near the positive trace for each integration in each file. Finally, we generate a normalized "white-light" light curve for four wavelength bins between 4 and 11 $\mu$m by summing the spectrophotometry in each bin and normalizing by the first integration. The light curves are displayed in Figure \ref{fig:var}, where we detect achromatic variability of approximately 25\% on timescales of about 1 minute in each band.

\subsection{LCO and Faulkes Telescopes}
\label{sec:LCO}
Regular optical observations of A0620 have been carried out for almost two decades with the 2-m Faulkes Telescopes (at Haleakala Observatory, Maui, Hawai`i, USA and Siding Spring Observatory, Australia), as part of a long-term monitoring program of $\sim 50$ low-mass XRBs coordinated by the Faulkes Telescope Project \citep{Lewis2008,Lewis2018}. The long-term monitoring uses the Bessel $V$ and $R$, and SDSS ${i}^{\prime}$ filters, shown in Figure \ref{fig:modulation} as cyan, red and gray symbols. 
\par
For the JWST campaign, we also made additional observations with the Las Cumbres Observatory \citep[LCO;][]{Brown2013} 1-m network (which includes telescopes in Australia, Chile, South Africa, Spain and USA), with additional images in Bessel $B$, $V$, $R$, SDSS ${i}^{\prime}$ and $z_{\rm s}$ and Pan-STARRS $Y$ filters. Within $\pm 2.5$ d of the JWST epoch, data were taken at UTC 2024-03-15 20:15--20:26, 2024-03-16 18:56--19:06 and 21:37--21:47, 2024-03-18 18:05--18:23 and 21:22--21:32, and 2024-03-19 09:21--09:32. The images were then processed using the ``X-ray Binary New Early Warning System (XB-NEWS)'', a real-time data analysis pipeline \citep[e.g.][]{Russell2019,Goodwin2020}. The XB-NEWS pipeline performs standard data reduction then several quality control steps to ensure that only good quality images are analyzed, then computes an astrometric solution for each image using Gaia DR2 positions\footnote{\url{https://www.cosmos.esa.int/web/gaia/dr2}}, performs aperture photometry of all the stars in each image, solves for zero-point calibrations between epochs \citep{Bramich2012}, and flux calibrates the photometry using the ATLAS All-Sky Stellar Reference Catalog \citep[ATLAS-REFCAT2;][]{Tonry2018}. XB-NEWS then performs multi-aperture photometry \citep[azimuthally-averaged PSF profile fitting photometry;][]{Stetson1990} on the target in each reduced image. Magnitude errors larger than $\sim 0.25$ mag are considered as marginal detections and are not included. The LCO data taken within 2.5 days of the JWST observations are shown in Figure~\ref{fig:sed} as filled green circles. The phase-folded data are highlighted in Figure ~\ref{fig:modulation} as filled blue triangles (V band), filled magenta circles (R band) and filled black circles ($i'$ band).   
Fluxes were de-reddened adopting a color excess $E(B-V) = 0.30\pm0.05$ \citep{Dincer2018}, which converts to $A_V = 0.93$, using $A_V = 3.1~ E(B-V)$.

\subsection{REM}
\label{sec:REM}

We observed A0620 with the 60cm Robotic Eye Mount (REM; \citealt{Zerbi2001,Covino2004}) telescope at La Silla Observatory (Chile). 
Observations took place between UTC 2024-03-17 03:00:40--03:45:19. We used the REMIR instrument equipped with the 2MASS H-band filter (1.662 $\mu m$).
We obtained twelve sets of five 30-second exposures, which were averaged in groups of five to subtract the sky background and improve the signal-to-noise ratio.
Aperture photometry is performed using \texttt{daophot} \citep{Stetson1987}, with an aperture radius of 6 pixels. The images were flux-calibrated against a selection of stars in the field from the 2MASS Point Source Catalogue\footnote{\url{https://irsa.ipac.caltech.edu/applications/2MASS/IM/}}.
The same color excess used in the optical analysis was adopted for de-reddening (Sec. \ref{sec:LCO}). REM data are shown as filled orange circles in Figure~\ref{fig:sed} and some of the green pluses (on the right side) in Figure~\ref{fig:modulation}. 
%
\subsection{HAWK-I}

We acquired high time-resolution, near-IR observations of A0620-00 in the H band (1.65 $\mu$m) using the HAWK-I instrument \citep{pirard2004}, installed on UT4 (Yepun) at the ESO VLT in Cerro Paranal, Chile. The dataset was obtained between UTC 2024-03-17 00:25--02:56 (program 110.23UL.002), employing the Fast Photometry configuration. In this mode, the readout is restricted to 16 adjacent windows of 128$\times$64 pixels per detector quadrant, achieving a temporal sampling of 1 second. The data are recorded in a series of cubes, each containing 94 frames, with brief ($\sim$3 s) inter-cube gaps due to buffer readout constraints. The pointing is adjusted so that both the science target and a nearby bright reference star (2MASS 06224288-0022059, H = 11.017$\pm$0.023) fall within the same quadrant (Q4). 
\par
Reduction of the raw data is carried out using the ULTRACAM software suite \citep{Dhillon2007}, where the reference star serves to guide aperture placement and photometric calibration. The centroid of the target is tracked relative to the reference star across each exposure, and the light curve is normalized using their flux ratio to correct for seeing-induced fluctuations. The reference star and an additional comparison source are confirmed to exhibit minimal variability. All frame timestamps are converted to the Barycentric Dynamical Time scale using the DE405 JPL ephemerides, following the prescription in \citet{eastman2010}. From the calibrated light curve, we estimate an average $H$-band magnitude of 14.81$\pm$0.02, corresponding to a flux density of approximately 1.23$\pm$ 0.01 mJy (not corrected for extinction). 
The phase-folded HAWK-I data are shown in Figure ~\ref{fig:modulation} in green.

\begin{table*}[]
\caption{Results of the VLT/FORS2 ($BVRI$ filters) polarimetric campaign. }         
\def\arraystretch{1.2}
\label{tab:pol}      
\centering                       
\begin{tabular}{c |c |c |c |c| c |c |c}       \hline   \hline           
 \multicolumn{2}{c|}{$B$} & \multicolumn{2}{c|}{$V$} & \multicolumn{2}{c|}{$R$} & \multicolumn{2}{c}{$I$} \\   
 \hline
 $P$ & $\theta$ & $P$ & $\theta$ & $P$ & $\theta$ & $P$ & $\theta$ \\
 \hline
 $1.46^{+0.38}_{-0.37} $ & $ 163.76^{+7.26}_{-7.27}$ & $0.60^{+0.17}_{-0.18} $ & $162.66^{+8.29}_{-8.30} $ & $0.96 \pm 0.04 $ & $162.86 \pm 1.19 $ & $0.80 ^{+0.08}_{-0.08} $ & $160.68^{+2.97}_{-3.08} $ \\
\hline\hline
\end{tabular}
\tablecomments{All of the polarization levels and angles are corrected for instrumental polarization. The interstellar polarization has also been subtracted, by means of a group of reference stars in the field. Uncertainties are quoted at the $\pm1 \sigma$ level. $R$-band polarization is the weighted mean of the 10 single measurements after averaging the asymmetrical errors.}

\end{table*}
\begin{figure}
    \centering
    \includegraphics[width=0.9\linewidth]{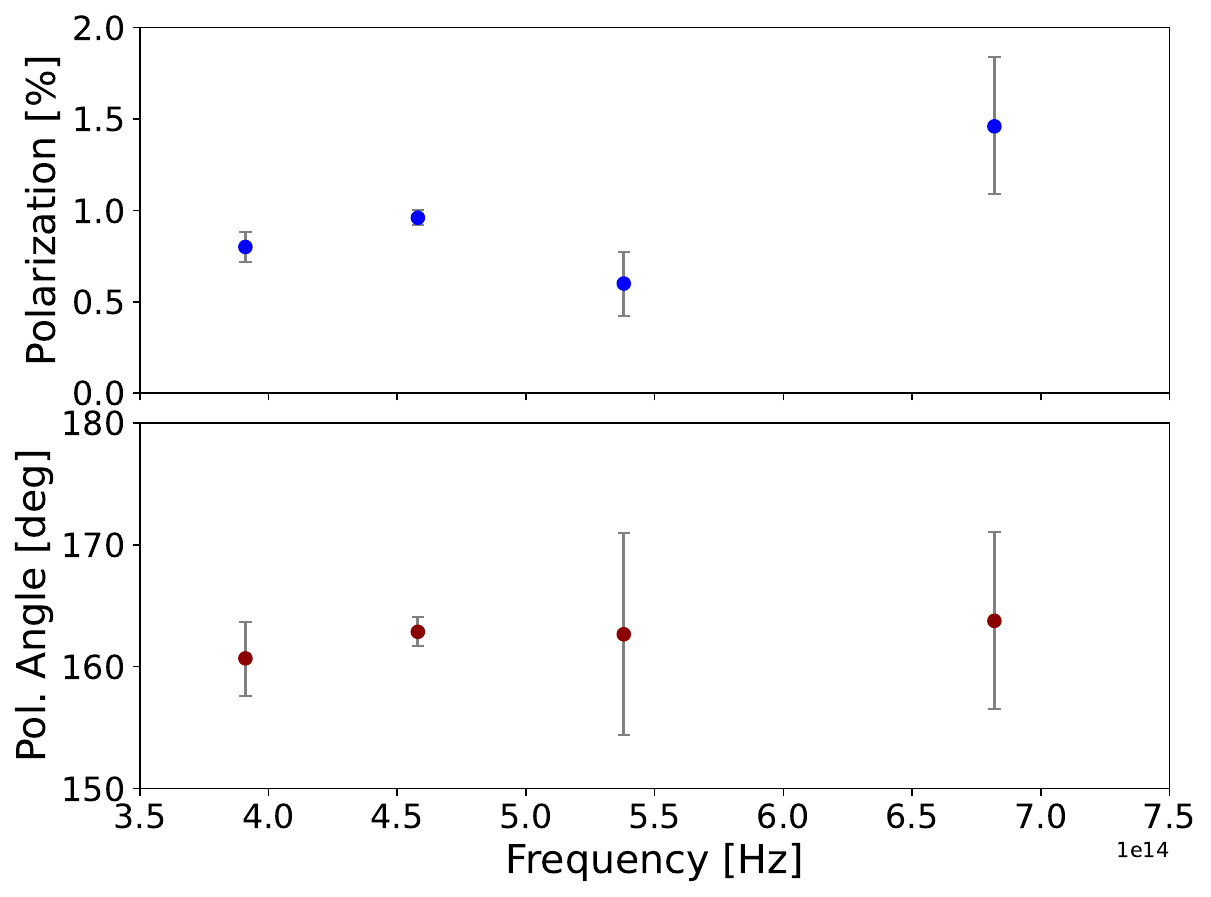}
    \caption{Results from the VLT polarimetry. Shown are the polarization level (top panel) and angle (bottom panel) as a function of frequency. All values are corrected for instrumental and interstellar polarization effects, and are listed in Table \ref{tab:pol}.}
    \label{fig:pol_SED}
\end{figure}

\subsection{VLT Polarimetry}
\label{sec:pol}
We conducted optical polarimetric observations of A0620 using the FOcal Reducer and low dispersion Spectrograph 2 \citep[FORS2;][]{appenzeller98}, mounted on the Very Large Telescope (VLT) at Paranal Observatory (Chile). Data were acquired in polarimetric mode using four optical filters: $b_{HIGH+113}$ ($B$; $\lambda_c = 440$ nm), $v_{HIGH+114}$ ($V$; $\lambda_c = 557$ nm), $R_{SPECIAL+76}$ ($R$; $\lambda_c = 655$ nm), and $I_{BESS+77}$ ($I$; $\lambda_c = 768$ nm). The polarimetric setup includes a Wollaston prism to split the incoming light into two orthogonally polarized beams—ordinary and extraordinary—and a mask to prevent beam overlap on the CCD. A rotating half-wave plate (HWP) enables the acquisition of images at four polarization angles, defined as $\Phi_i = 22.5^{\circ}(i-1)$ for $i = 1, 2, 3, 4$. Each polarization measurement consists of a set of four images taken at the four HWP angles.
\par
The target was initially observed in the $R$-band only, with 10 image sets acquired, each comprising four 20-second exposures at the four HWP angles. These observations took place between UTC 2024-03-19 00:10:12.776 and UTC 2024-03-19 00:53:07.845. Later that night, between UTC 2024-03-19 01:02:05.013 and 2024-03-19 01:16:01.854, one set of polarimetric observations was taken in each of the $B$, $V$, and $I$ filters, with exposure times of 25 seconds in $V$ and 35 seconds in both $B$ and $I$.
Raw images are calibrated via bias subtraction and flat-field correction. Aperture photometry is performed using \texttt{daophot} \citep{Stetson1987}, adopting a 5-pixel radius aperture, corresponding to approximately $0.7''$.
To measure the linear polarization of A0620-00, we employ the methodology outlined in \citet{baglio20} and references therein. The procedure starts by computing the parameter $S(\Phi)$ at each half-wave plate (HWP) angle, defined as:

\begin{equation}
S(\Phi)=\left( \frac{f^{o}(\Phi)/f^e(\Phi)}{f^o_u(\Phi)/f^e_u(\Phi)}-1\right)\bigg/\left( \frac{f^{o}(\Phi)/f^e(\Phi)}{f^o_u(\Phi)/f^e_u(\Phi)}+1\right),
\end{equation}

where $f^{o}(\Phi)$ and $f^{e}(\Phi)$ represent the ordinary and extraordinary fluxes of the target, while $f^{o}_u(\Phi)$ and $f^{e}_u(\Phi)$ correspond to the same quantities for an unpolarized comparison star in the field. This normalized ratio yields a polarization signal that is largely independent of instrumental and atmospheric effects.
The $S(\Phi)$ parameter is related to the degree of polarization ($P$) and polarization angle ($\theta$) through the equation:

\begin{equation}\label{eq_cos}
S(\Phi) = P\, \cos 2(\theta - \Phi),
\end{equation}

which allows us to extract $P$ and $\theta$ by fitting the measured $S$ values across the four HWP angles. To improve the robustness of this fit, we include six field stars in each epoch under the assumption that they are intrinsically unpolarized. This approach effectively compensates for any instrumental polarization\footnote{FORS2 has consistently shown stable and minimal instrumental polarization (below $0.3\%$) across all filters over the past ten years, based on regular observations of unpolarized standard stars.} and, to first order, also corrects for interstellar polarization along the line of sight, assuming all stars share a similar dust column. This assumption is further reinforced by the selection of reference stars in the field, which exhibit comparable parallax measurements in the Gaia DR2 archive, indicating that they are likely subject to a similar level of interstellar polarization.
\par
Following \citet{baglio20}, we determine the best-fit values of $P$ and $\theta$ by maximizing a Gaussian likelihood using the Nelder–Mead optimization algorithm \citep{gao12}, and sample the posterior distributions via a Markov Chain Monte Carlo (MCMC) method based on the affine-invariant ensemble sampler \citep{Foreman-Mackeyetal2013, Hogg&Foreman2018}. Each MCMC chain is initialized with small Gaussian perturbations around the maximum likelihood estimates, with the first third of each chain discarded as burn-in. Convergence is assessed following the stability criteria in \citet{Sharma2017}, and fit quality is evaluated according to \citet{Lucy2016}.
\par
To improve the accuracy of the polarization angle, we apply a calibration correction derived from observations of the polarized standard star Hiltner 652, obtained on 2024 April 2 using the same instrumental setup. Due to saturation effects, calibration is possible only in the $B$ and $V$ bands, where the required corrections were minimal ($1.91^{\circ}$ in $B$ and $2.26^{\circ}$ in $V$). Since the tabulated polarization angle of Hiltner 652 is the same within errors across the $BVRI$ filters \citep{Cikota2017}, we adopt the average of the $B$ and $V$ corrections ($2.08^{\circ}$) and applied it also to the $R$ and $I$ bands.

The results of the polarimetric analysis in the four filters are summarized in Table \ref{tab:pol} and shown in Fig. \ref{fig:pol_SED}.
Both the polarization level and angle are consistent with previous studies \citep{Russell2016,Kravtsov2024}. However, unlike \citet{Kravtsov2024}, we do not observe any clear evidence of a rotation of the polarization angle with frequency. It is worth noting that our error bars are larger or comparable to the variation they report, and our observations were taken at a single orbital phase, whereas their results were averaged over approximately one full orbital period.

\subsection{MeerKAT}\label{sec:meerkat}

A0620 was observed by the MeerKAT radio telescope \citep{meerkat} using the L-band (856-1712 MHz) receivers during the JWST campaign, from UTC 2024-03-17 from 14:04:55 to 18:51:44. These observations were made as part of the X-KAT program (MeerKAT Proposal ID: SCI-20230907-RF-01). The observing block included eight 30-minute scans of A0620 each sandwiched by 2-minute scans of a secondary (phase and delay) calibrator, J0632+1022. Additionally, J0408-6545 was observed at the start and end of the observing block for 5 minutes as a primary (bandpass, flux, leakage) calibrator. Finally, J0521+1638, was observed for 10 minutes at the start of the observing block as a polarization angle (cross hand phase and delay) calibrator. Observations were made with a dump time of 8 seconds and with 4096 frequency channels. The total integration time on A0620 was 4 hours. The observational data were reduced using the semi-automated pipeline {\tt polkat} \citep{polkat}. The various software packages used in this pipeline were accessed using {\tt singularity} for software containerization \citep{singularity}. The data were processed while in the Measurement Set (MS) format \citep{measurement_set}. First generation iterative RFI flagging and calibration was done using the Common Astronomy Software Applications ({\tt CASA}; \citealt{casateam}) software (v6.5.0). The target field was flagged further using {\tt tricolour} \citep{tricolour} and then imaged using {\tt wsclean} \citep{wsclean}. A deconvolution mask was made using the mask making tool {\tt breizorro} \citep{breizorro} before phase and delay self calibration were performed using {\tt CASA}. For all imaging, joined frequency and joined polarization deconvolution, using 8 channels and a Briggs weighting of $-0.3$ \citep{briggs}, was used.

A0620 was not detected at the position of the JWST source, and we place a $3\sigma_\text{rms}$ upper limit of $13.5\:\mu\text{Jy}$ on the flux density from A0620 at 1.28~GHz, represented by the downward green triangle in Figure~\ref{fig:sed}. 
\subsection{VLA}
\label{sec:vla}
VLA observations were taken in C configuration (maximum baseline 3.4 km), starting at UTC 2024-03-17 20:56:57 (12.7 hours after the start of JWST observations), for five hours. We observed in C (4--8 GHz) and X (8--12 GHz) bands using the 3-bit correlators at central frequencies of 6.2 and 9.8 GHz, respectively, obtaining nearly 110 minutes on the source in each observing band.  For both bands, we used J0542+4951 as the primary flux density and bandpass calibrator, and we used J0641$-$0320 as our secondary phase calibrator to solve for the time-dependent complex gains.  Data were reduced following standard procedures {\tt CASA} (v6.5.0).  We started with the pipeline-calibrated products supplied by NRAO, to which we applied a small amount of additional flagging.  Images were produced in each observing band using the {\tt CASA} task {\tt tclean}, using two Taylor terms to account for the wide fractional bandwidth, and adopting Briggs weighting with ${\tt robust=1.5}$ to balance sensitivity with minimizing sidelobes from nearby sources. A0620 was not detected in either image at the position of the JWST source, down to 3$\sigma_{\rm rms}$ limits of $<$11.4 $\mu$Jy (6.2 GHz) and $<$13.5 $\mu$Jy (9.8 GHz).  We, therefore, stacked the C- and X-band observations to create a slightly deeper image, which revealed a marginal (2.8 $\sigma$) source consistent with the known location of A0620, which we measured to have a flux density of 8.5$\pm$3.0 $\mu$Jy at 8.0 GHz (measured by fitting a two-dimensional Gaussian using the task {\tt imfit}, with the major and minor axes fixed to the size of synthesized beam, $5.1$ and $3.6$ arcsec, respectively).

\subsection{Stellar-Subtracted Spectrum}
\label{sec:modeling}

Interpreting the spectra of hard and quiescent XRBs at MIR wavelengths hinges on careful modeling of the systems' optical and near-IR emission, from which the donor star contribution can be teased out. 
In the absence of variability and/or contamination, the donor star contribution can be modeled using a suitable stellar template, normalized to the de-reddened optical and near-IR fluxes (see, e.g., \citealt{Muno2006}). However, decades of optical and near-IR monitoring of A0620 have established the presence of a well-understood variability pattern that needs to be accounted for. 
Analysis of optical and near-IR data from 1981 to 2007 by \citet{Cherepashchuk2019} show that, since the decline from its 1975 outburst, A0620 has oscillated between two types of quiescent states of varying durations: passive and active, with intermediate loops state (see also \citealt{Hynes2003,Zurita2003,Neilson2008,Cantrell2008,Cantrell2010,Shugarov2016,vanGrunsven2017,Dincer2018}).
Each state exhibits different levels of activity. In the passive state, the phase-folded light curve displays regular sinusoidal modulations, with little or no flickering variability. This behavior is attributed to ellipsoidal variations caused by changes in the geometrical cross-section of the tidally distorted donor star, which fills its Roche lobe.  
In contrast, the active state shows increased brightness and fluctuations compared to the baseline ellipsoidal modulation, with variations on very short timescales. 
\begin{figure}
    \centering
    \includegraphics[width=\linewidth]{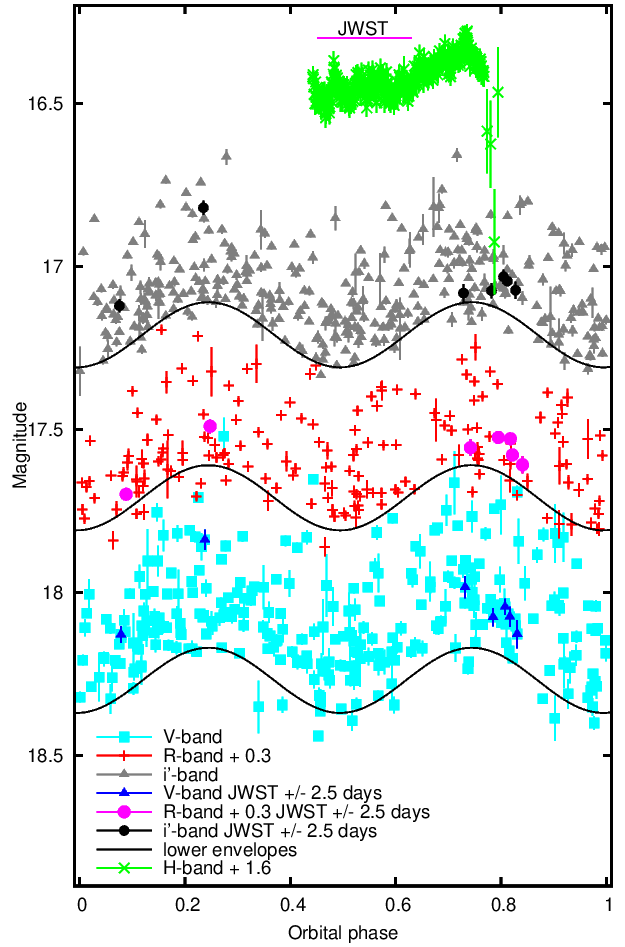}
    \caption{Two decades of optical monitoring of A0620 with the Faulkes Telescope. In addition to the phase-folded historical data, the new LCO data taken within $\pm2.5$ days of the March 2024 JWST observations are shown as filled blue triangles (V band), filled red circles (R band), filled black circles ($i'$ band). Green crosses show the VLT/HAWK-I (left) and REM (right) near-IR data (H band). The magenta horizontal line at the top brackets the phase interval corresponding to the JWST observations. The solid black curves are the lower envelopes of each filter; they represent the intrinsic emission from the donor star in each band \citep[e.g.][]{Cantrell2008,Cantrell2010}. }
    \label{fig:modulation}
\end{figure}
\par
Figure~\ref{fig:modulation} summarizes two decades of optical monitoring of A0620 with the Faulkes Telescopes, in V, R, and $i$' bands (cyan, red, and gray points). The light curves are folded on the orbital period of 7.7523377 hr using the updated ephemeris of \cite{Cherepashchuk2019}; $T_0 = {\rm JD}~2457332.601507$. Phase zero is the upper conjunction of the black hole (when the optical star is in front of the black hole). The baseline ellipsoidal modulations traced by the passive state are shown as solid black curves \cite[derived similarly to][]{Cantrell2010,Bernardini2016,Baglio2022}, and referred to as lower envelopes. The phase-folded March 2024 LCO measurements of A0620 (blue, magenta, and black) are superimposed to the historical Faulkes data. In each band, the 2024 data lie systematically above the lower envelope, meaning that, at the time of the observations, A0620 was in an (at least slightly) active state. 
As a result, if we were to normalize a stellar template to the measured fluxes, we would significantly overestimate the donor star contribution. 
\par
Following e.g., \cite{Cantrell2010} and \cite{Dincer2018}, we estimate the average V-band magnitude associated with the lower envelope curve at the time (i.e., phase) of each JWST observation (the phase interval covered by the JWST observations is illustrated by the magenta horizontal bar at the top of Figure~\ref{fig:modulation}), and use that to normalize the amplitude of a BT-Setll stellar template \citep[][and references therein]{Allard2013} with photospheric temperature $T=4,500$K, surface gravity $\log g=4.5$ and solar abundance \footnote{Obtained from \url{https://svo2.cab.inta-csic.es/theory/newov2/index.php?models=bt-settl}.}. The spectral type of the donor star has generally been categorized within the K3V to K5V range, with the latter preferred by \cite{Cherepashchuk2019} (however, see \citealt{Zheng2022}, who argue for a K2V type). By comparison with similar stellar types, we estimate that the uncertainty in the stellar type leads to a 10\% systematic flux error in the template normalization.
\par 
Uncontaminated stellar fluxes at the appropriate phase are then subtracted from the measured values. The stellar-subtracted MIRI LRS spectrum and photometric data are shown in Figure~\ref{fig:fit}. 
%

\begin{figure}
 \centering{
    \includegraphics[width = 1\columnwidth]{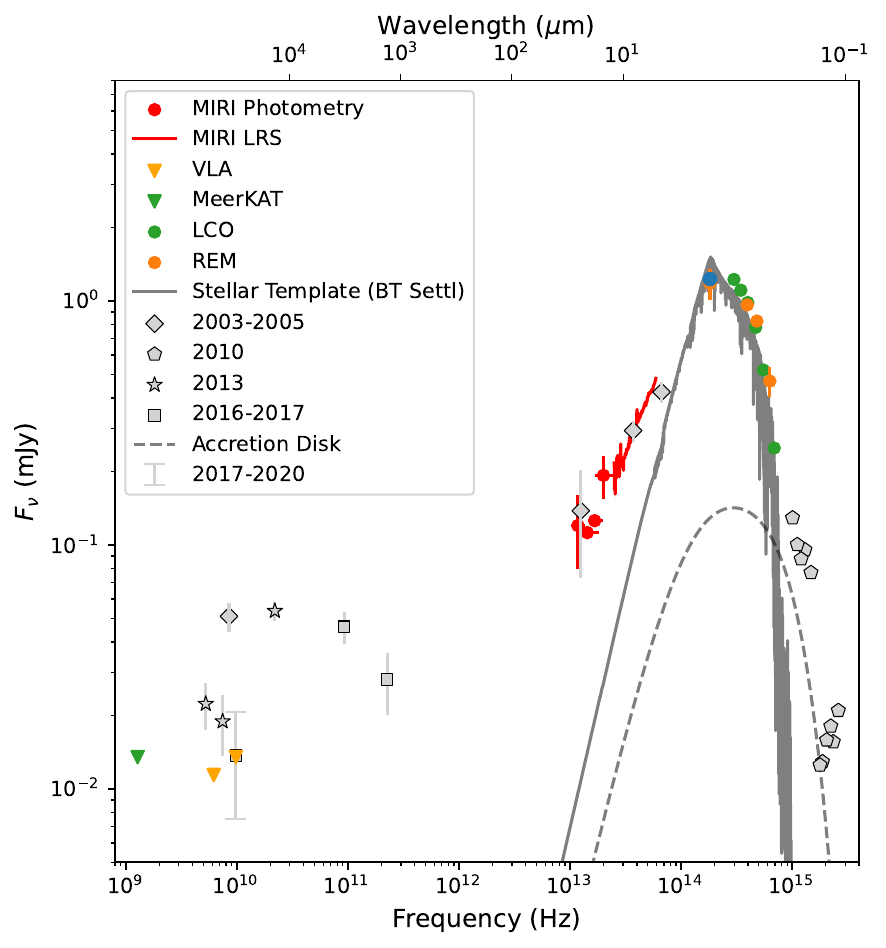}\hspace{1cm}
        }
    \caption{Radio to optical SED of A0620, including new (MeerKAT, VLA, JWST, VLT, REM, and LCO, from March 2024) and archival data. Data are de-reddened if applicable. 
    The solid dark gray curve traces the BT Settl stellar template for the donor star, normalized to the V-band flux of the lower envelope for the historical data at the time/phase of the JWST observations, as described in \S~\ref{sec:modeling}. The dashed gray curve corresponds to a multi-temperature blackbody disk with inner and outer radius of $10^4$ and $1.8 \times 10^5$ gravitational radii, and a mass accretion rate of $2.6\times 10^{16}$ g s$^{-1}$. Sources of archival data: diamonds are from \citet{Gallo2007}; pentagons: \citet{Froning2011}; stars: \citet{Dincer2018}; squares: \citet{Gallo2019}; light gray vertical bar:  \citet{dePolo2022}.}
    \label{fig:sed}
\end{figure}

\section{Modeling the MIR Excess}
\label{sec:mir}
Multiple investigations of A0620, conducted over the last two decades, have significantly advanced our understanding of the system's behavior across all wavelengths. The near-IR band in particular is crucial for determining the system's mass function. As discussed in the previous Section, there is strong agreement that a significant portion of the near-IR (and optical) flux during active states is non-stellar in origin. The physical origin of this non-stellar emission is uncertain. 
The nature of the accretion flow in A0620 (and in highly sub-Eddington black hole X-ray binaries in general) is a subject of active debate. Since its decline from the 1975 outburst, A0620 has been emitting at an X-ray luminosity that is between eight and nine orders of magnitude below the Eddington limit, with a factor of 10 variability in X-ray emission on timescales of years \citep{Dincer2018}. At these low luminosities, the flow can no longer maintain the conditions necessary for an optically thick, geometrically thin disk that extends to the innermost regions. Instead, it is generally accepted that the inner disk transitions to an optically thin, hot, and radiatively inefficient flow, where most of the liberated accretion power is either advected inwards, as in advection-dominated accretion flows (ADAFs; cf. \citealt{Narayan1995}), or lost to an outflow, as described in adiabatic inflow-outflow solutions (ADIOS; cf. \citealt{Blandford1999,Blandford2004}). Early attempts to fit the quiescent SED of A0620 with an ADAF model approximate the overall spectral shape and the measured near-UV spectrum reasonably well, though they consistently and significantly overestimate the system's optical-IR flux and spectral shape (e.g., figures 6 and 7 in \citealt{Narayan1997}).  
\par
The presence of an outer, optically thick accretion disk is confirmed by the detection of double-peaked H$\alpha$ emission lines in A0620 \citep{Froning2011}. Several authors have attempted to determine the fractional contribution of the accretion disk relative to that of a relativistic jet or the hot flow at near-infrared and optical wavelengths. \cite{Dincer2018} find that the variability of the non-stellar emission increases with wavelength, whereas the variability of the total emission is lower and achromatic. This is consistent with an increasing jet contribution at longer wavelengths. A jet contribution is also consistent with the linear polarization excess measured by \cite{Russell2016}, as well as the alignment of the magnetic field vector with the axis of the radio jet, which was resolved during the system's 1975 outburst (see, however, \citealt{Kravtsov2024}). \cite{Dincer2018} also separate the non-stellar emission into a non-variable component and a flaring component. The spectral shape of the non-variable component appears to be flat, albeit with significant excess in the J and H bands, which may be associated with the peak of the hot accretion flow spectrum. 
\par
\cite{Cherepashchuk2019} conclude that, during the active state, the non-stellar component contributes to as much as 50\% of the near-IR flux. Incidentally, these authors ascribe the transitions between the active and passive state of A0620 to mass transfer enhancements through the inner Lagrangian point, which are likely driven by non-stationary processes in the donor star's atmosphere. 
In a more recent study, based on strictly simultaneous photometric and spectroscopic data, \cite{Zheng2022} concluded that the outer accretion disk contributes most of the non-ellipsoidal variations in the optical band.  Adding to the controversy, \cite{Muno2006} argue that the excess, as measured by Spitzer, could be interpreted as arising from blackbody emission from a dusty disk of circumbinary material.
\par 
This complex pattern of behavior argues for multiple components contributing to the optical and near-IR flux at any given time. By extension, multiple emission sources could be responsible for the MIR excess first detected by Spitzer and later confirmed by the JWST MIRI data presented in this work. \\

The stellar-subtracted MIRI spectrum of A0620, as shown in Figure~\ref{fig:fit}, is of far superior quality compared to the Spitzer data. The extrapolation of the MIRI spectrum at GHz frequencies is clearly well below the VLA/MeerKAT limits. 
This indicates that the decision to include or discard the marginal detection with the VLA significantly impacts the modeling: if only the radio upper limits are taken into account, they exert minimal constraining power on the SED, which can be reasonably well modeled by a single component/functional form. In contrast, if the marginal VLA detection resulting from the stacking analysis is regarded as valid, it necessitates the consideration of a two-component model. We examine these scenarios separately below.

\subsection{Bayesian Inference: Single Component Modeling}
\label{subsec:bayes}
We start by considering two simple functional forms: a power-law and a blackbody spectrum.  We exclude the wavelength ranges between $7-7.8~\mu\mathrm{m}$ and $10.2-10.8~\mu\mathrm{m}$ to avoid potential contamination from emission lines, which will be discussed separately (\S~\ref{sec:LRSfitting}).  The radio upper limits are formally included in the fit.
For each data point $k$ of flux density $F_k$, the model assumes $F_k\sim\mathcal{N}(M_k, \sigma_k^2)$, where $\mathcal{N}$ denotes a normal distribution, $M_k$ is the model flux density at the corresponding wavelength and $\sigma_k$ is the flux density uncertainty. 
\par
The corresponding log-likelihood is
\begin{align}
\ln\mathcal{L}_k=-\frac{1}{2}\ln(2\pi)-\ln(\sigma_k)-\frac{1}{2}\left(\frac{M_k-F_k}{\sigma_k}\right)^2. 
\end{align}
The first term will be used to calculate the model evidence later. 
\par
\par The log-likelihood for the upper limits, denoted as $F_{\mathrm{lim},k}$, is:  

\begin{multline}
\ln\mathcal{L}_k = \ln P(\mathcal{N}(M_k, \sigma_k^2) < F_{\mathrm{lim},k}) \\
= \ln \left( \frac{1}{2} \left[ 1 + \mathrm{erf} \left( \frac{F_{\mathrm{lim},k} - M_k}{\sqrt{2} \sigma_k} \right) \right] \right).
\end{multline}
In this case, we set $\sigma_k=0.01F_{\mathrm{lim},k}$ such that any value of $M_k > F_{\mathrm{lim},k}$ would be highly penalized, and the exact choice of $\sigma_k$ has little impact on the fitting, as long as it is much lower than $F_{\mathrm{lim},k}$. The joint log-likelihood function for all data points is then: 
\begin{align}
    \ln \mathcal{L} = \sum_{k\in\text{Detections}} \ln \mathcal{L}_k + \sum_{j\in\text{Upper Lmits}} \ln \mathcal{L}_j.
\end{align}

\par The flux density for the blackbody and the power-law models are: 
\begin{equation}
\begin{cases}
    M_{\mathrm{BB}}=\left(\frac{R}{D}\right)^2\pi B(\nu, T) \\
    M_{\mathrm{PL}}= C \nu^\alpha
\end{cases}
\end{equation}
Here, $R$ is the blackbody radius, $D$ is the distance to A0620, $B(\nu, T)$ is the Planck function, $C$ is the power-law normalization constant, and $\alpha$ is the power-law spectral index. 
The prior distributions are given in Table \ref{tab:fit}.\par

\begin{figure*}
 \centering{
    \includegraphics[width = 1.6\columnwidth]{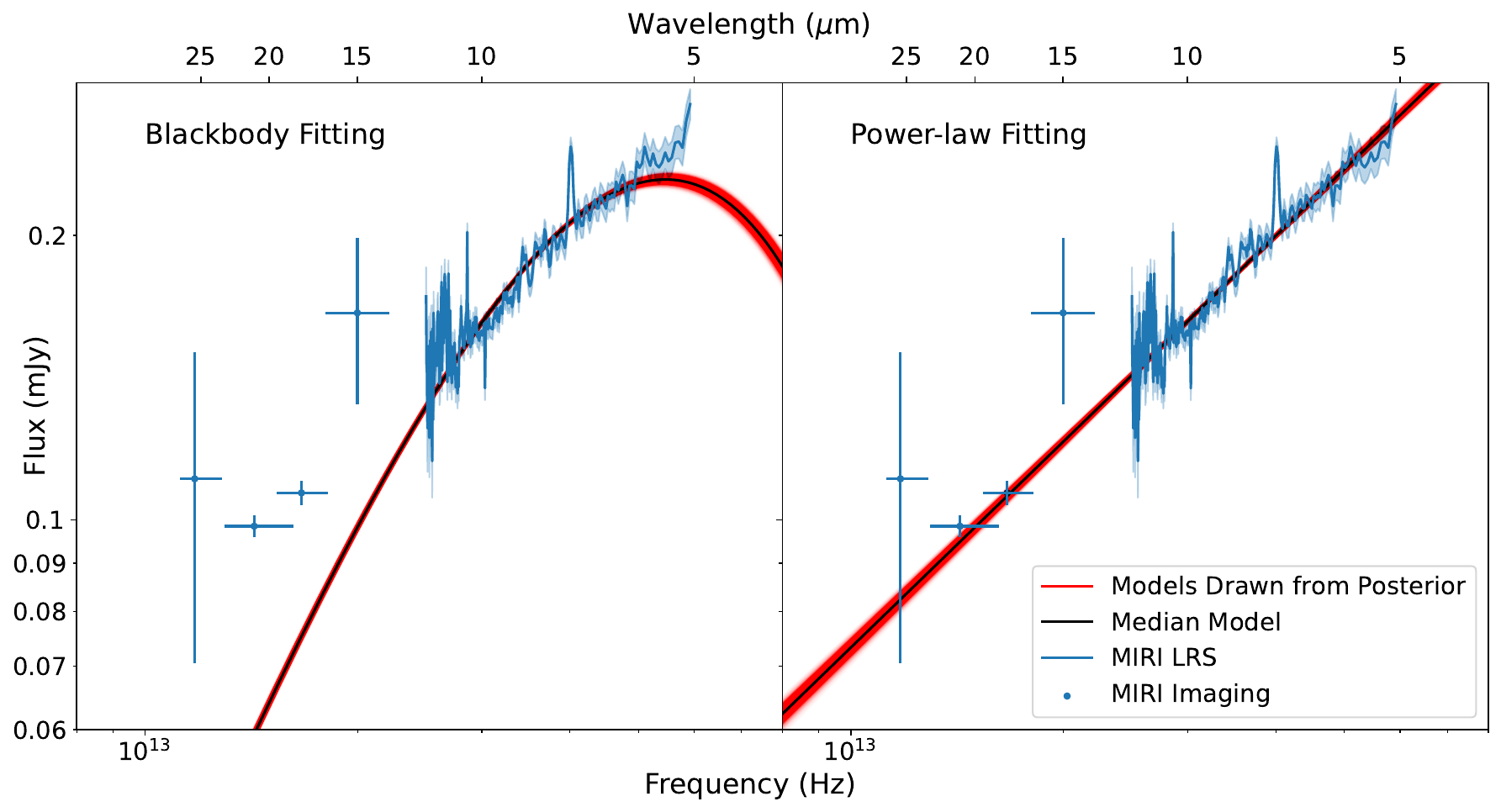}\hspace{1cm}
        }
    \caption{Stellar-subtracted MIRI data, fitted with a blackbody model (left) and a power-law model (right). The 15 $\mu$m data point errors account for the measured 40\% variability. The red curves are drawn from the posterior distributions, with the solid black curve tracing the median model. A power law of the form $F_{\nu}\propto \nu^{0.72\pm 0.01}$ is strongly favored by the data (see Table \ref{tab:fit}).   }
    \label{fig:fit}
\end{figure*}

Sampling is conducted with \texttt{DynamicHMC.jl},\footnote{\url{https://www.tamaspapp.eu/DynamicHMC.jl/stable/}} which implements Hamiltonian Monte Carlo \citep{Betancourt17}. The fitting process is initialized with a set of optimal solutions obtained with \texttt{Optim.jl}. 
The fitted parameters for both the blackbody and power-law models are listed in Table \ref{tab:fit}. The resulting models are illustrated in Figure~\ref{fig:fit}, where the red curves represent 4,000 models randomly drawn from the posterior distribution, and the black curves trace the median models.
\par
Visual inspection of Figure~\ref{fig:fit} suggests that the power-law model can fit the data reasonably well, while the blackbody model significantly underestimates the photometry at long wavelengths. 
\par For a quantitative assessment, we calculate the models' Bayes factor, i.e., the ratio of their model evidence. The model evidence is defined as $\int\mathcal{L}(\theta)\pi(\theta)d\theta$, where $\mathcal{L}(\theta)$ is the likelihood, $\pi(\theta)$ is the prior, and $\theta$ represents all the parameters. Although directly calculating model evidence is generally challenging and often requires specialized or simplified methods, our models are relatively simple, allowing us to perform the integration numerically. The Bayes factor comparing the power-law model to the blackbody model is $10^{107}$. Since a Bayes factor of 100 is considered decisive evidence favoring one model over another \citep{Kass95}, we conclude that the data strongly favor the power-law model. The posterior median of the slope is $0.72 \pm 0.01$ (see Table~\ref{tab:fit} for a full list of parameters). 
\par
A second, compelling line of evidence against blackbody emission is the $\sim$1-minute timescale variability seen by JWST, which is much shorter than any timescales that would be associated with a circumbinary disk (dynamical, thermal, and viscous). 
\par
The inferred spectral slope is also much shallower than the $\propto \nu^2$ dependence expected from the Rayleigh-Jeans tail of a multi-temperature accretion disk. 
In light of these considerations, we conclude that a power law model is strongly favored by the data, and do not explore the possibility of fitting the stellar-subtracted MIR spectrum of A0620 with a multi-temperature blackbody model.  
%
\begin{table}
\caption{Stellar-Subtracted Spectral Modeling}
\def\arraystretch{1.2}
\label{tab:fit}
\vspace{0.3cm}
\begin{tabular}{lcccc}
\hline
\hline
 &  \multicolumn{2}{c}{Blackbody} \\
& $(R/D)\times 10^{11}$ & $T$ (K)  \\
\hline
  Priors & $\mathcal{U}(0,1000)$ & $\mathcal{U}(0,\, 5\times 10^{5})$  \\
  Results & $6.93^{+0.06}_{-0.06}$ & $929^{+7}_{-7}$  \\
  \hline
  &  \multicolumn{2}{c}{Power-law }\\
  & $\log (C/\mathrm{mJy)}$ & $\alpha$\\
  \hline
  Priors  & $\mathcal{U}(-10,10)$& $\mathcal{U}(-100,100)$\\
  Results & $-0.41^{+0.01}_{-0.01}$ & $0.72^{+0.01}_{-0.01}$\\
\hline
\hline

\end{tabular}
\end{table}

\subsection{Multiple Component Modeling}
\label{subsec:bayes2}
Considering the marginal detection at 8.0 GHz reported in Section~\ref{sec:vla} complicates the modeling. Acceptable fits to the radio-MIR SED with a single model can be achieved only if the LRS error bars are scaled by a factor $s = \sqrt{N_{\rm spec}/N_{\rm phot}}$, where \( N_{\rm spec} = 240 \) and \( N_{\rm phot} = 5 \) represent the number of spectral and photometric data points, respectively\footnote{Without proper adjustments to the LRS errors, the radio data point would effectively lose its constraining power. Adopting a scale factor to weaken the constraining power of the spectral data sets is a common procedure when attempting to fit broadband SEDs for galaxies; see, e.g., \cite{Lopez2016}.}. 
Doing so still yields a somewhat less inverted power law spectrum ($\alpha=0.60\pm 0.02$); a single blackbody still fails to reproduce the MIR excess. 
However, this procedure effectively means reducing the constraining power of the LRS spectrum to that of a single photometric point, or equivalently reverting to the 20 year old Spitzer data. 
\par 
If the marginal radio detection is indeed associated with synchrotron radiation from a partially self-absorbed jet, sensible modeling requires two separate components that are highly degenerate, as the jet's spectral shape and break frequency are entirely unconstrained. To cover all possible extremes, a `maximal jet model' is one where the jet emission smoothly connects the measured flux densities at 6.5 GHz and 25 $\mu$m, yielding a power law of the form $F_{\nu} \propto \nu^{+0.34}$. If this power further extends to 5 $\mu$m, that would represent close to 40\% of the flux at that wavelength. We stress that this scenario is completely arbitrary and should only be considered as a firm upper limit to the jet contribution to the MIR. It will not be further considered in the remainder of this Paper.

\subsection{LRS Line Modeling}
\label{sec:LRSfitting}
The LRS spectrum shown in Figure~\ref{fig:fit} exhibits a strong emission feature at approximately $7.5~\mu\mathrm{m}$. Interestingly, the same feature is evident in the LRS spectrum of the black hole XRB GRS~1915+105 presented by \citet{Gandhi2025}. We follow their methodology to investigate its origin. 
\par
First, we fit the LRS spectrum with a fourth-order polynomial curve to accurately model the continuum around the emission feature. As above, we exclude the wavelength ranges between $7-7.8~\mu\mathrm{m}$ and $10.2-10.8~\mu\mathrm{m}$ to avoid potential contamination from less significant nearby emission lines. We utilize the \texttt{NEBULAR} package \citep{Schirmer16} to generate narrow emission line spectra, assuming a helium abundance of 10\% by volume. The model spectra are convolved with wavelength-dependent Gaussian kernels to approximate the finite spectral resolution power of MIRI LRS (\citealt{Kendrew15}; the spectral resolution of the LRS at 7.5 $\mu$m is $R\approx$ 90, corresponding to a velocity resolution of $\approx 3,000$ km s$^{-1}$). 
\par
Figure~\ref{fig:specline} illustrates our modeling results. The $7.5~\mu\mathrm{m}$ emission feature can be explained by the blend of hydrogen recombination lines: H(6-5) at $7.46~\mu\mathrm{m}$, H(8-6) at $7.50~\mu\mathrm{m}$, and H(11-7) at $7.51~\mu\mathrm{m}$, contributing 75\%, 20\%, and 5\% of the observed line flux, respectively. The equivalent width of the emission feature (before subtracting the stellar spectrum) is $EW\approx 0.016$ $\mu$m. 
\par
The figure includes two representative model spectra at different temperatures, demonstrating that slight temperature variations significantly affect the ionization state of helium. 
Temperatures in excess of 20,000 K are excluded by the absence of prominent \iona{He}{ii} emission lines. Albeit not formally a fit, these recombination models are consistent with a gas temperature of 12,000 K. We conclude that the MIRI LRS spectrum of A0620 is broadly consistent with H recombination lines from gas with neutral helium. We will return to the interpretation and physical origin of the line in \S~\ref{sec:discussion} below. 
%
\section{Discussion}
\label{sec:discussion}

This work presents observations of the prototypical quiescent black hole X-ray binary A0620 (\lx$\approx 10^{-9} L_{\rm Edd}$) obtained using the JWST MIRI in March 2024. These observations are complemented by a comprehensive ground-based campaign with contributions from MeerKAT, the VLA, HAWK-I, VLT, REM, and LCO.
The system’s MIR spectrum reveals significant excess emission above the Rayleigh-Jeans tail of the donor star, confirming earlier results obtained with Spitzer (Figure~\ref{fig:sed}). After subtracting the donor star contribution, the 5-25 $\mu$m spectrum of A0620 is well-characterized by a power law of the form $F_{\nu} \propto \nu^{0.72 \pm 0.01}$ (errors do not account for systematics).
\par
A radio counterpart is only marginally detected, with a flux density of $8.5\pm 3.0$ $\mu$Jy at 8.0 GHz. Even if a faint radio jet were present, its emission cannot account for the observed MIR spectrum. Assuming an extreme scenario, where the jet's partially self-absorbed synchrotron spectrum connects the inferred flux density at 8.0 GHz with that at 25 $\mu$m, yields an $F_{\nu} \propto \nu^{0.34}$ dependence, which underestimates the MIR excess at shorter wavelengths.
Indirect evidence against the (speculative) "maximal jet" scenario presented above is provided by the VLT polarimetry data (\S~\ref{sec:pol}). These data appear consistent with polarization resulting from scattering within the disk. A similar conclusion is drawn by \citealt{Kravtsov2024}, based on a recent, similar dataset. \\

\par
The MIR spectral shape and rapid variability—a 40\% flux flare at $15~\mu$m, along with 25\% achromatic variability in the LRS data—strongly rule out a circumbinary disk as the source of the excess MIR emission in A0620. The detection of a prominent emission feature at 7.5 $\mu$m, resulting from the blending of three hydrogen lines, indicates the presence of baryonic material with a temperature of $\simlt 20,000$ K.
Determining the origin of the emission lines at 7.5 $\mu$m is critical for the interpretation of this data set. We note that the same spectral feature was identified in the LRS spectrum of the high-luminosity low-mass black hole XRB GRS~1915+105 \citep{Gandhi2025}. In this case, the MIR continuum is attributed to thermal bremsstrahlung from an obscuring wind, along with hydrogen recombination lines. Furthermore, even though the authors do not model the emission line spectrum in the range blue-ward of 14 $\mu$m, the same feature is apparent in the Spitzer/IRS spectrum of the high-mass black hole XRB Cygnus X-1 \citep{Rahoui2011}. In this system, the continuum is interpreted as arising from the sum of the (O-type supergiant) donor star Rayleigh-Jeans tail plus bremsstrahlung emission from the stellar wind\footnote{The SED of a spherically symmetric stellar wind with a constant mass loss rate ($\dot{M}$) and velocity ($v_{\infty}$) leads to a spectrum of the form $F_{\nu} \propto (\frac{\dot M}{v_{\infty}})^{4/3}\nu^{2/3}$ in the radio-IR \citep{Wright1975}. The intermediate slope between that of optically thin homogeneous plasma ($\propto \nu^{-0.1}$) and optically thick plasma ($\propto \nu^{2}$) occurs because, at any given frequency, the emission is observed down to the depth where the wind becomes optically thick.}. Both systems emit X-rays at more than 5 orders of magnitude higher luminosities; in all cases, the feature was detected during spectrally hard X-ray states.\\

Preliminary analysis (\S~\ref{sec:LRSfitting}) suggests that the emitting gas is likely cooler than 20,000 K. The first critical question to address is whether the 7.5 $\mu$m line complex could originate from the outer accretion disk, given that the low spectral resolution of the LRS instrument does not allow for the resolution of double-peaked features. Basic modeling dismisses this possibility. Figure \ref{fig:sed} presents a multi-color disk blackbody (dashed line) from a Shakura-Sunyaev accretion disk, extending out to the circularization radius of A0620 (0.017 AU, or approximately $1.8\times 10^5$ gravitational radii), with an accretion rate of $2.3 \times 10^{16}$ g s$^{-1}$ (\citealt{Froning2011}). The inner radius is set to a few $10^4$ gravitational radii so as not to exceed the far-UV flux measured by STIS in 2010. The estimated disk flux at 7.5 $\mu$m is about 30 $\mu$Jy. If the 7.5 $\mu$m emission feature were produced by the disk, it would imply a nonphysically high equivalent width, exceeding that of H$\alpha$, which is known to have a factor of approximately 50 higher yield.
\par
The lines must be produced by a population of excited ions in warm gas--sufficiently hot to ionize hydrogen and allow recombination but not so hot as to excite helium (\S~\ref{sec:LRSfitting}). This is inconsistent with the extreme temperatures expected at the base of a relativistic jet/compact corona ($10^{8-10}$ K; \citealt{Markoff2015} and references therein) and in an ADAF, where the ions can be up to 100 times hotter than the leptons (\citealt{YuanNarayan2014} and references therein). Similarly to the case of Cygnus X-1 and GRS~1915+105, we argue that the most likely explanation for both the MIR continuum of A0620 and the observed recombination lines is bremsstrahlung emission from an outflow, consistent with, e.g., two-zone ADIOS models \citep{Blandford1999, Blandford2004, Begelman2012}. 
\par
\begin{figure}[t!]
\centering
\hspace{1cm}
\includegraphics[width=\linewidth]{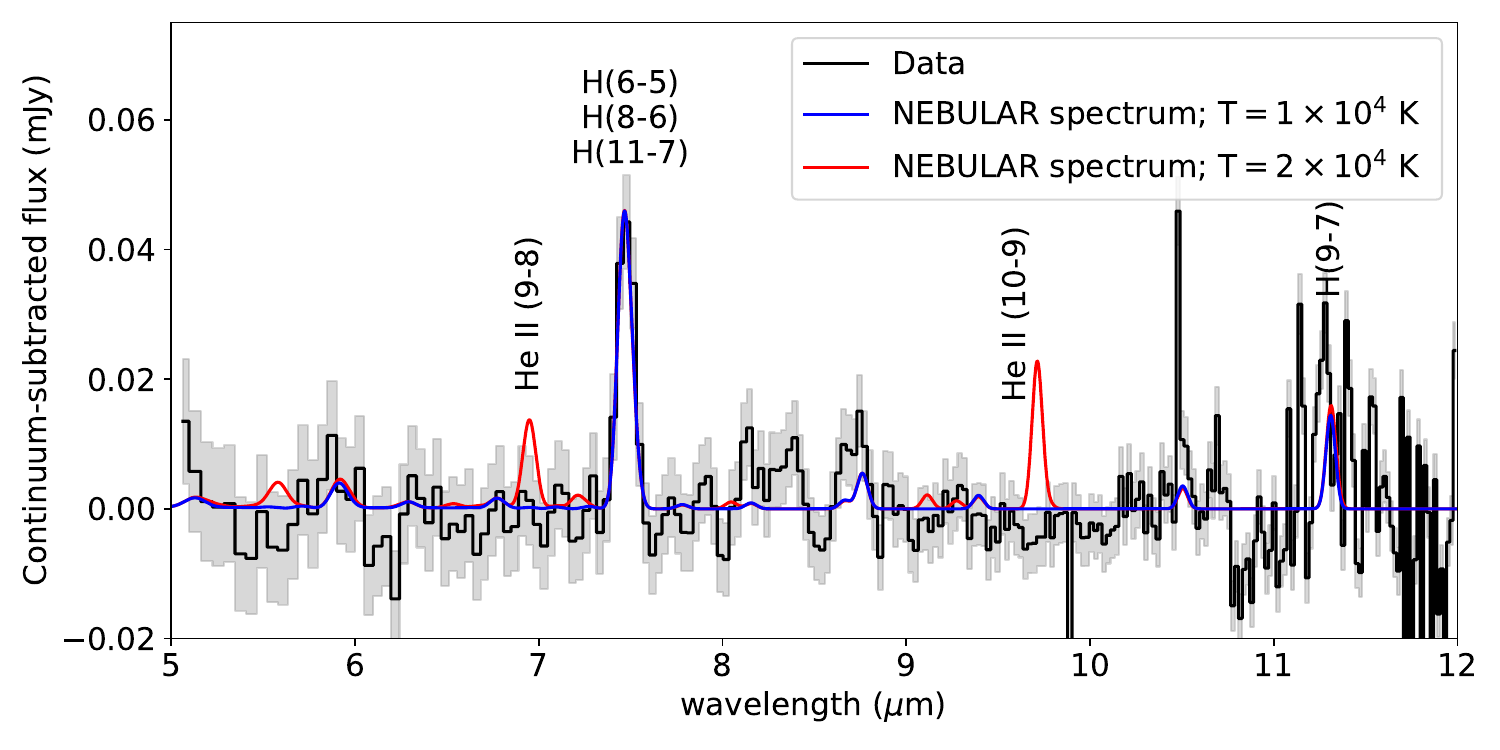}
\caption{Fit the $\sim 7.5$ micron line seen by MIRI LRS, using the {\tt NEBULAR} package. This feature arises from the blend of the H(6-5), H(8-6), and H(11-7) recombination lines. The spectrum is highly sensitive to the assumed temperature; the lack of strongly ionized helium line indicates that the gas temperature does not exceed 20,000 K.   }
\label{fig:specline}
\end{figure}

While no model SEDs are available in the literature for ADIOS, \texttt{CLOUDY} \citep{Ferland2013} can be utilized for a consistency check. We simulate an AGN-like spectrum where the so-called blue bump is replaced with a 4,500 K blackbody spectrum, which represents the donor star contribution. The bolometric luminosity is a few $10^{32}$ erg s$^{-1}$, with 10\% in the form of X-ray luminosity. The X-ray photon index is assumed to be $\Gamma=2$ \citep{Dincer2018}. This spectrum illuminates a homogeneous cloud located at a distance between $10^{4-5}$ gravitational radii for a 6.6 solar mass black hole. The transmitted spectrum reproduces the hydrogen recombination lines observed in the LRS spectrum for number densities in the range $n\approx 10^{10-11}$ cm$^{-3}$, which is similar to the estimated range of outflow densities in GRS~1915+105 (see Section 4.5 of \citealt{Gandhi2025}). The inferred temperature range is somewhat higher than that estimated using the \texttt{NEBULAR} package, i.e., between 30,000-40,000 K. Even though this temperature exceeds the ionization energy for helium, the \texttt{CLOUDY} spectrum, similarly to the \texttt{NEBULAR} spectrum shown in Figure~\ref{fig:specline}, does not exhibit prominent helium lines at 9.7 $\mu$m.
\par
This exercise confirms that the observed emission feature at 7.5 $\mu$m is {\it qualitatively}\footnote{Since  \texttt{CLOUDY} does not calculate the intrinsic emission from the gas cloud, we can not claim any rigorous quantitative agreement with the measured spectrum.} consistent with arising from the recombination of hydrogen within warm gas that is photo-ionized by a source whose SED resembles the observed shape and luminosity of A0620 in quiescence. The same gas is bound to emit thermal bremsstrahlung radiation, even though we cannot rule out additional contributions from synchrotron radiation, e.g., from a relativistic jet (\citealt{Markoff2005, Markoff2015, Malzac2013, Lucchini2022} and references therein) or a hot, radiatively inefficient inflow (\citealt{Veledina2013, YuanNarayan2014, PoutanenVeledina2014} and references therein; see \cite{Liska2024} for state-of-the-art radiative two-temperature general relativistic magneto-hydrodynamic simulations, including quiescent black hole XRBs). Deeper radio observations, ideally conducted simultaneously with MIR observations, may help distinguish between the possible explanations discussed above.
\par
A quantitative estimate of the total radiative output from such an outflow from a theoretical standpoint is beyond the scope of this work. However, we note that the expected continuum is likely to differ from the $\nu^{2/3}$ dependence expected from a spherical wind (Begelman, private communication\footnote{Specifically, an ADIOS wind likely introduces an additional power-law dependence on frequency compared to the spherical case, due to the fact that the $(\frac{\dot{M}}{v_{\infty}})$ ratio is a function of radius.}).
\par
From the wind number density, we can derive an order of magnitude estimate for the wind mass loss rate, $\dot{M}=4 \pi r^2  n m_H v_{w}$, where $r$ is the wind launching radius, $v_{w}$ its velocity, and $m_{H}$ is the hydrogen atom mass (we are implicitly assuming a homogeneous wind, with a covering factor of the order of unity). An upper limit to the wind velocity can be derived from the lack of measurable shifts or broadening, allowing us to set $v_{w} \simlt 3,000$ km s$^{-1}$ (i.e., the LRS resolution at 8 $\mu$m). Assuming a launching radius between $10^{4-5}$ gravitational radii (or $\approx$$5 \times 10^{9-10}$ cm), this yields mass loss rates in the range $\simlt 1.5 \times 10^{15-16}$ [$1.5 \times 10^{14-15}$] g s$^{-1}$ for $n=10^{11}$ [$10^{10}$] cm$^{-3}$. Despite the large uncertainties, these values correspond to a few percent up to more than 50\% of the accretion rate estimated at the outer edge of the accretion disk, i.e., $2.4 \times 10^{16}$ g s$^{-1}$, as derived by \cite{Froning2011} based on the inferred luminosity of the hotspot (equation 1 in their paper).
Interestingly, the corresponding dynamical timescale of the wind ($t_\mathrm{dyn} = r/v_w$) would be $\simgt 15-170$ s, which is in excellent agreement with the observed variability timescale at 15 $mu$m.
\par
The MIR energetics are also interesting. The luminosity of the stellar subtracted MIR continuum at 8 $\mu$m is on the order of a few $10^{31}$ erg s$^{-1}$, or ten times higher than the average X-ray luminosity of the system \citep{Dincer2018}.  This rules out the possibility that the MIR variability results from the reprocessing of minute-scale X-ray variability. Overall, the data presented in this work provide observational support to the notion that the highly sub-Eddington luminosities of quiescent black hole XRBs arise from the fact that a large fraction of the mass supply at large radii is actually lost to a wind, rather than being advected through the black hole event horizon.

\begin{acknowledgements}
This work is based on observations made with the NASA/ESA/CSA James Webb Space Telescope. The data were obtained from the Mikulski Archive for Space Telescopes at the Space Telescope Science Institute, which is operated by the Association of Universities for Research in Astronomy, Inc., under NASA contract NAS 5-03127 for JWST. Support for GO Program number 03832 (EG and ZZ) was
provided through a grant from the STScI under NASA contract NAS5-03127.
JM is supported by an NSF Astronomy and Astrophysics Postdoctoral Fellowship under award AST-2401752.
AV acknowledges support from Academy of Finland grant 355672.
DMR is supported by Tamkeen under the NYU Abu Dhabi Research Institute grant CASS. MCB acknowledges support from the INAF-Astrofit fellowship. RMP acknowledges support from NASA under award no. 80NSSC23M0104. Nordita is supported in part by NordForsk. We are grateful to Mitch Begelman for helpful feedback to an earlier version of this manuscript. 
\end{acknowledgements}

\bibliographystyle{aasjournal}
\bibliography{bibliography_XRB.bib}

\end{document}